\documentclass[fleqn,usenatbib]{mnras}

\newcommand{\ktwo}{\emph{K2}}
\newcommand{\jwst}{\emph{JWST}}

\newcommand{\mjup}{M$_{\rm JUP}~$}

\newcommand{\teff}{\ensuremath{T_{\rm eff}}}

\newcommand{\water}{H$_2$O}
\newcommand{\methane}{CH$_4$}

\usepackage[T1]{fontenc}
\usepackage{ae,aecompl}
\usepackage{tablefootnote}

\usepackage{graphicx}	
\usepackage{amsmath}	
\usepackage{amssymb}	

\usepackage{newtxtext,newtxmath}
\usepackage{lineno,multirow}

\title[The Atmosphere of a $\lesssim$5 Myr-old Exoplanet]{CO, H$_2$O, and CH$_4$ in the Dusty Atmosphere of a $\lesssim$5 Myr-old Exoplanet}
\author[Gaidos \& Hirano]{Eric Gaidos$^{1,2}$ 
\thanks{E-mail: gaidos@hawaii.edu.}
and Teruyuki Hirano$^{3,4}$
\\
$^{1}$Department of Earth Sciences, University of Hawai'i at M\={a}noa, 1680 East-West Rd, Honolulu, HI  96822, USA\\
$^{2}$Institute for Astrophysics, University of Vienna, T\"{u}rkenschanzstrasse 17, 1180 Vienna, Austria\\
$^{3}$Astrobiology Center, 2-21-1 Osawa, Mitaka, Tokyo 181-8588, Japan\\
$^{4}$National Astronomical Observatory of Japan, NINS, 2-21-1 Osawa, Mitaka, Tokyo 181-8588, Japan\\
}

\date{}

\pubyear{}

\hypersetup{draft}
\begin{document}
\label{firstpage}
\pagerange{\pageref{firstpage}--\pageref{lastpage}}
\maketitle

\begin{abstract}
Very young massive planets are sufficiently luminous by their internal heat of formation to permit detailed studies, including spectroscopy of their atmospheres with large telescopes at sufficient resolution ($\lambda / \Delta \lambda \gtrsim 1000$) to identify major constituents to inform models of planet formation and early evolution.  We obtained 1--2.4\micron\ ($YJHK$) spectra of the planetary-mass ``b" companion of 2MASS~J04372171+2651014, a 1-3 Myr-old M dwarf member of the Taurus star-forming region, and one of the youngest such objects discovered to date.  These indicate the presence of CO and possibly \water\ and \methane\ in the atmosphere, all suggesting a \teff\ of around 1200K, characteristic of a L-T transition spectral type and consistent with previous estimates based on its luminosity and age.  The absence or attenuation of spectral features at shorter wavelengths suggests the presence of micron-size dust, consistent with the object's red color.  The spectrum of 2M0437b resembles those of the HR 8799 planets, especially the innermost ``b" planet, with the exception of a pronounced flux deficit in the $H$-band of uncertain origin. 
\end{abstract}




\section{Introduction}
\label{sec:intro}

One central avenue to understanding the formation and early evolution of planets is to directly observe them around members of star-forming regions and young stellar associations/clusters.  Until recently, only Jupiter-sized planets could be detected in emission although the advent of \jwst\ and Extremely Large Telescopes should bring smaller planets within reach \citep{Currie2022}.   Observations of giant planets can test formation scenarios (i.e., core accretion vs. disk fragmentation) and thermodynamic conditions (i.e., ``hot start" vs. ``cold start") \citep{Spiegel2012,Marleau2014,Wallace2021}.  The envelopes of giant planets represent samples of natal disk gas containing clues to disk evolution \citep{Cridland2019,Zhang2020,Notsu2020}.  The youngest planets could still host detectable circumplanetary disks \citep{Zhu2016,Wu2020,ChenX2022}, the properties of which would be indicative of their formation.   A description of giant planet demographics informs studies of protoplanetary disk evolution, since these objects could be responsible for the formation of inner disk cavities \citep{vanderMarel2018} and warps \citep{Nealon2018}.  

At the distances of the nearest star-forming regions ($\sim$100 pc), resolution of planets and substellar companions at separations 10-100 au ($\sim$0".1-1") with ground-based instruments requires adaptive optics (AO) systems to partially recover diffraction-limited resolution.  The advent of AO systems on large telescopes has led to a steady stream of discoveries \citep{Currie2022}, but high-order wavefront correction requires a bright guide star, and the members of young clusters ($\lesssim$ few Myr) are distant and often extincted by interstellar dust and hence faint at optical wavelengths.  This limitation has excluded many low-mass members -- the dominant population -- from earlier surveys.  This omission is impactful, since for the same contrast ratio it is possible to detect older/less massive/closer planets around lower mass, less luminous stars. 

One approach is to perform wavefront correction at longer (near-infrared) wavelengths where cooler/reddened stars are brighter, while another is to use an artificial laser guide star (LGS) as the source to complement lowest-order ("tip-tilt") correction with a natural guide star.  We used LGS-AO to detect a faint ($\Delta K \approx 7$) object 0".9 from the mid-M dwarf member of the 1-5 Myr-old Taurus star-forming region 2MASS~J04372171+2651014 \citep[hereafter 2M0437,][]{Gaidos2022b}.  Nearly three years of astrometric monitoring confirmed that the companion is co-moving with its host star; a comparison with stellar models indicated that the host star is 2-5 Myr old, and a comparison of the planet's luminosity with exoplanet models with that age range suggests a mass of 3-5 \mjup.  2M0437b is one of the youngest, lowest mass companions discovered to date and thus an important benchmark in exoplanet research.  While the object's luminosity and color suggest \teff\ is 1400--1500K, characteristic of late L-type dwarfs, the $H$-$K$ color of this object, like many other low-gravity (planetary-mass) companions, is significantly redder than late L dwarfs.  This discrepancy has been explained by the suspension of dust in the low-gravity, convective atmospheres of these objects \citep{Liu2016}.  

Photometry contains limited information, but a near-infrared spectrum of 2M0437b can be compared to models to refine the effective temperature \teff, surface gravity, and atmospheric composition.  Spectra of late L-type dwarfs with this \teff\ contain deep absorption features by \water, CO, and possibly \methane.  At this temperature, emission will be modulated by clouds of refractory dust, settling of which is an indicator of the object's gravity and thus mass.  All else being equal, lower gravity planets are expected to be cloudier, redder, and have weaker absorption features. Also, ongoing accretion in a circumplanetary disk could manifest itself by emission in hydrogen lines, e.g., the Brackett $\gamma$ line (2.17$\mu$m).  However, the small separation between 2M0437b and its much brighter host and its scattered light poses a challenge to conventional spectroscopy.

Here we present near-infrared spectra of 2M0437b obtained with two instruments and AO systems at ground-based observatories on Maunakea in Hawai'i.  One of these spectrographs has been previously used to detect CO, \methane, and \water\ in the $\sim$30 Myr-old giant planet HR 8799b \citep{Barman2015,Roche2018} and search for emission in H lines in disk-embedded planets \citep{Uyama2021}, while similar instruments have been used to observe substellar companions \citep{Wilcomb2020,Petrus2021,Cugno2021}.    

\section{Observations and Data Reduction}
\label{sec:observations}

Observations of the 2M0437 system were performed with the InfraRed Camera and Spectrograph \citep[IRCS,][]{Tokunaga1998,Kobayashi2000} and AO188 system \citep{Hayano2010} at the Subaru telescope on UT 11 August 2018, and the OH-Suppressing Infra-Red Imaging Spectrograph (OSIRIS) \citep{Larkin2006} and AO system on Keck-1 on UT 11 February 2020 and on 29 October 2021.  Low-spectral resolution ($\lambda/\Delta \lambda \approx 470$), 52 mas spatial scale, 0.95-1.5 \micron\ spectra were obtained with IRCS and a $zJH$ grism, $0\farcs15$ slit, and integration times of 180-540 sec.  Spectra of the A0 star HD 29526 were obtained for telluric correction, and Ar lamp exposures were used for wavelength calibration.  Spectral images were reduced and spectra extracted using custom software.  Apertures that were 8 and 21 pixels wide in the spatial dimension were used to extra spectra of source and sky, respectively.  

OSIRIS samples the focal plane with 1019 lenslets, each producing a spectrum with an average resolution of $\lambda / \Delta \lambda \approx 3800$.  The observing scheme consisted of multiple integrations on the target followed by a single sky integration offset 5" to the north.  Integrations were obtained with Hbb ($\lambda $= 1.473--1.803\micron) and Kbb ($\lambda = 1.965--2.381$ \micron) filters and the 0".035 lenslet scale, with a field of view of 0".56 $\times$ 2".24.  Observations of the A0-type star BD+27 716B were obtained for telluric correction.  Integration times were 600 sec for 2M0437 and 2 sec for BD+27 716B.  Weather for both nights was good, but problems with the AO optical bench made the data from the first run largely unusable and only the October data were analyzed here.   

\begin{figure}
 \includegraphics[width=0.5\columnwidth]{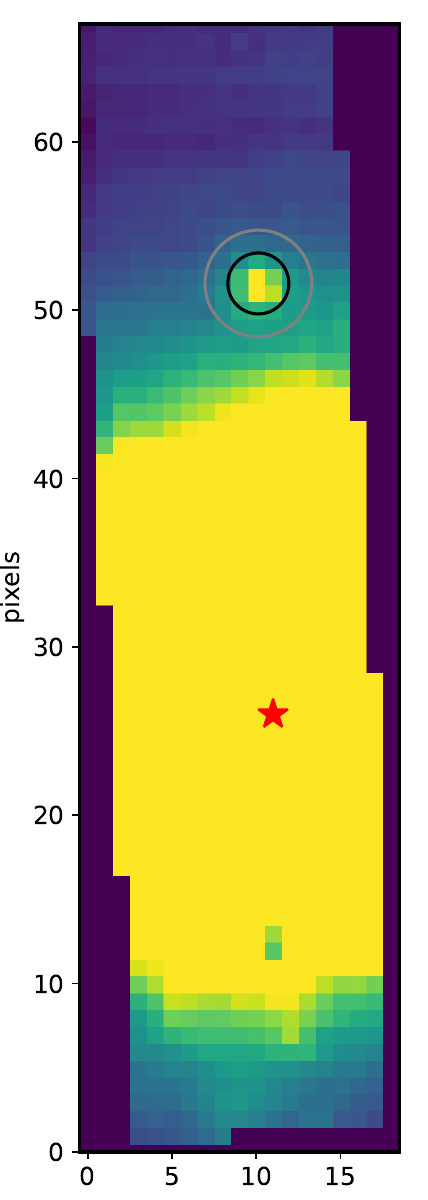}
	\includegraphics[width=0.5\columnwidth]{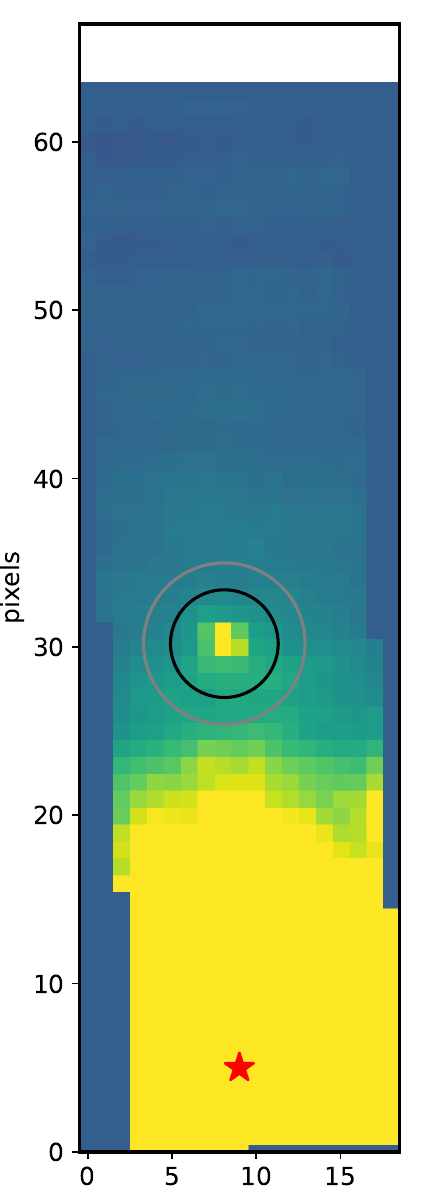}
    \caption{Single reduced 600-sec OSIRIS frames in $H$-band (left) and $K$-band (right) of the 2M0437 system.  Each wavelength-summed image is 0".56 by 2".24, and oriented with a PA $\approx$ 68 deg.   The host star is marked by the red star while the companion is marked by concentric circles; the inner circle is for aperture photometry while the background is computed between the inner and outer circles.  The colormap spans $\pm3\times$ the standard deviation in the background annulus.}  
    \label{fig:frame}
\end{figure}

The OSIRIS Data Reduction Pipeline \citep[DRP,][]{Lyke2017,Lockhart2019} was used to convert the 2-d images into 3-d datacubes.  Because contemporary rectification matrices were not available for Hbb and Kbb, matrices from January 2021 were used.  (Newer matrices were not obtained until October 2022).  1-d spectra of the A0 star for telluric correction were extracted using the {\tt telluricARP} DRF template, divided by a 9840K black-body, and hydrogen lines were excised.   The residual spectra were then used to correct the 2-d spectra of 2M0437 produced by the {\tt fullARP} DRF template.  Dark frames were not used since a sky frame was subtracted from each science frame.  Figure \ref{fig:frame} show single sky- and telluric-corrected Hbb and Kbb frames showing 2M0437b clearly resolved from its star.  Spectra of 2M0437b were extracted from 3-d data-cubes by first stacking all wavelength channels to create an integrated 2-d image, then fitting a Gaussian to the source in that image within a 2.5 pixel-radius aperture (black circle in Fig. \ref{fig:frame}), after subtracting the background determined as the median in an annulus of pixels between that aperture and a 4-pixel concentric circle (grey circle in Fig. \ref{fig:frame}).  The best-fit amplitude, full-width half-maximum (FWHM), and centroid coordinates were obtained and the process repeated once with the aperture re-centered and re-scaled to 2.5 times the $\sigma$ from the initial Gaussian fit.  The apertures were then fixed and a Gaussian fit was performed in each wavelength channel, holding the centroid and FWHM fixed, and solving for the amplitude.  The background in each channel was calculated more precisely by fitting a second-order surface to the background annuals and then interpolating within the source aperture.  In both cases, pixels deviating by more than 5 standard deviations (``robust" calculation) were excluded from these fits.  Spectra were computed for each individual 600-sec integration for a given pass-band and summed.  The summed spectra are plotted as blue points Figs. \ref{fig:HKspec}, with a running 21-channel median plotted as the blue line.  

OSIRIS spectra of 2M0437b corrected by A0 star spectra still contained significant telluric features, perhaps due to variation in airmass and water vapor content between the times and airmass of the observations.  We corrected the spectra further by a best fit of a model consisting of the deviations of the A0 star spectrum from unity, multiplied by a free parameter, times a simple third-order polynomial fit to the intrinsic 2M0437b spectrum.  The polynomial is only used to provide sufficient degrees-of-freedom to obtain an adequate fit to the tellurics; it is not removed from the spectrum.  The corrected spectra are plotted as the green lines in Fig. \ref{fig:HKspec}. 

Errors were estimated using 100 Monte Carlo (MC) simulations of the image in each wavelength channel.  Each MC realization was the actual image plus Gaussian-distributed noise, with the RMS of the noise set to the robust standard deviation ($\sqrt{\pi/2} \times$ the median absolute deviation) of the residuals of the fit to the background annulus.  As before, a Gaussian profile was fit to the image of the object, and the robust standard deviation of the 100 MC amplitudes was adopted as the uncertainty in the observed value.  These uncertainties are represented by the light green lines in Fig. \ref{fig:HKspec}.

We flux-calibrated the spectrum in each band-pass by convolving and integrating over the appropriate response function.  Since the pipeline produces spectra with units of counts sec$^{-1}$ per channel and the channels are linear with wavelength, these numbers are divided by wavelength to yield energy per unit wavelength.  We adopted the appropriate response functions and zero-point fluxes archived in the Filter Service of the Spanish Virtual Observatory (SVO).   We used the $K'$-magnitude of 17.21 derived from repeated Keck-2/NIRC2 observations in \citet{Gaidos2022b} and the $H$-magnitude of 18.05 from the Subaru IRCS discovery imaging reported in \citet{Gaidos2022b}.  Since the $H$-band imaging used pairs of saturated and unsaturated images, we checked this photometry by performing relative star-companion photometry in the OSIRIS data.  We convolved and integrated both the spectrum of the companion and the star (within an aperture the same size as that used for the companion to minimize Strehl effects) over the relevant response function, finding $\Delta H = 7.71$ with $\sim$0.06 mag consistency between individual integrations.  Combined with 2MASS photometry of the star, these yield $H=18.37$, 0.32 mags fainter than the values derived by \citet{Gaidos2022b}.  This could reflect systematics in either instrument, and/or intrinsic variability as found among some ultracool objects \citep[e.g.,][]{Bowler2020b}

\section{Results}
\label{sec:results}

\begin{figure}
    \centering
    \includegraphics[width=\columnwidth]{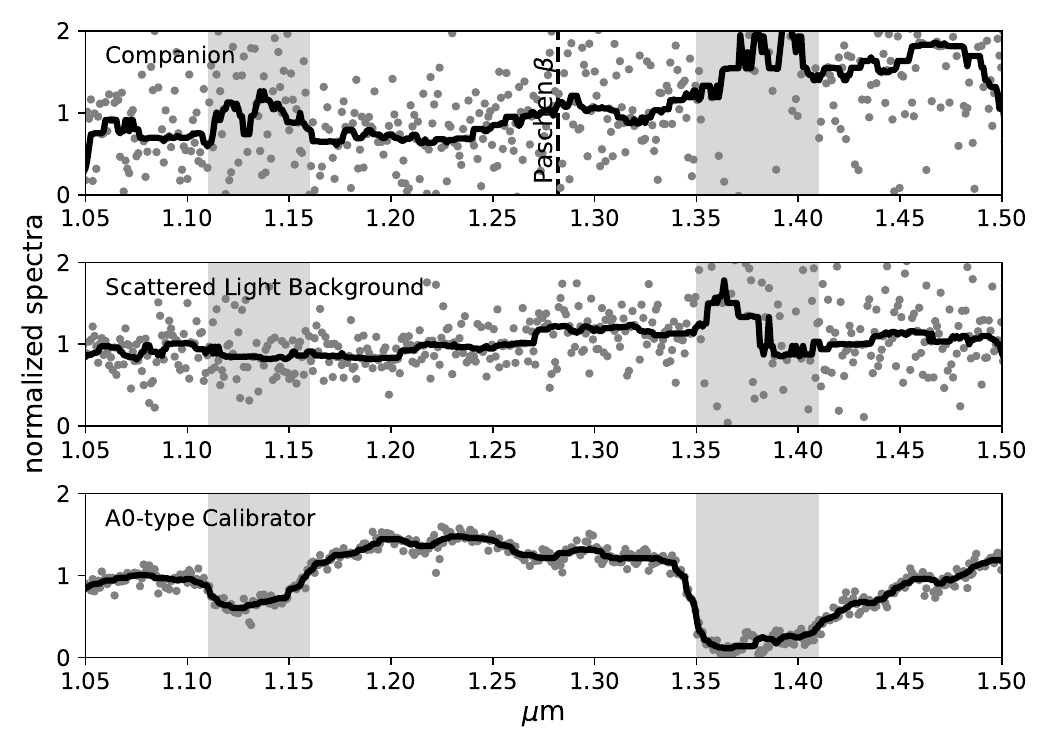}
    \caption{IRCS $YJ$-band spectrum of 2M0437b (top), the scattered light from the host star that was subtracted (middle), and the A0-type star HD 29526 used to correct telluric absorption.  The solid curves are 31-pixel (0.027 \micron) running medians (21 pixels = 0.019 \micron\ for the A0 star) and the grey areas denote regions between the Y, J, and H bands where the telluric correction is high and spectra are unreliable.  The location of the Paschen $\beta$ line of H is marked.}
    \label{fig:ircs}
\end{figure}

\begin{figure*}
	\includegraphics[width=\textwidth]{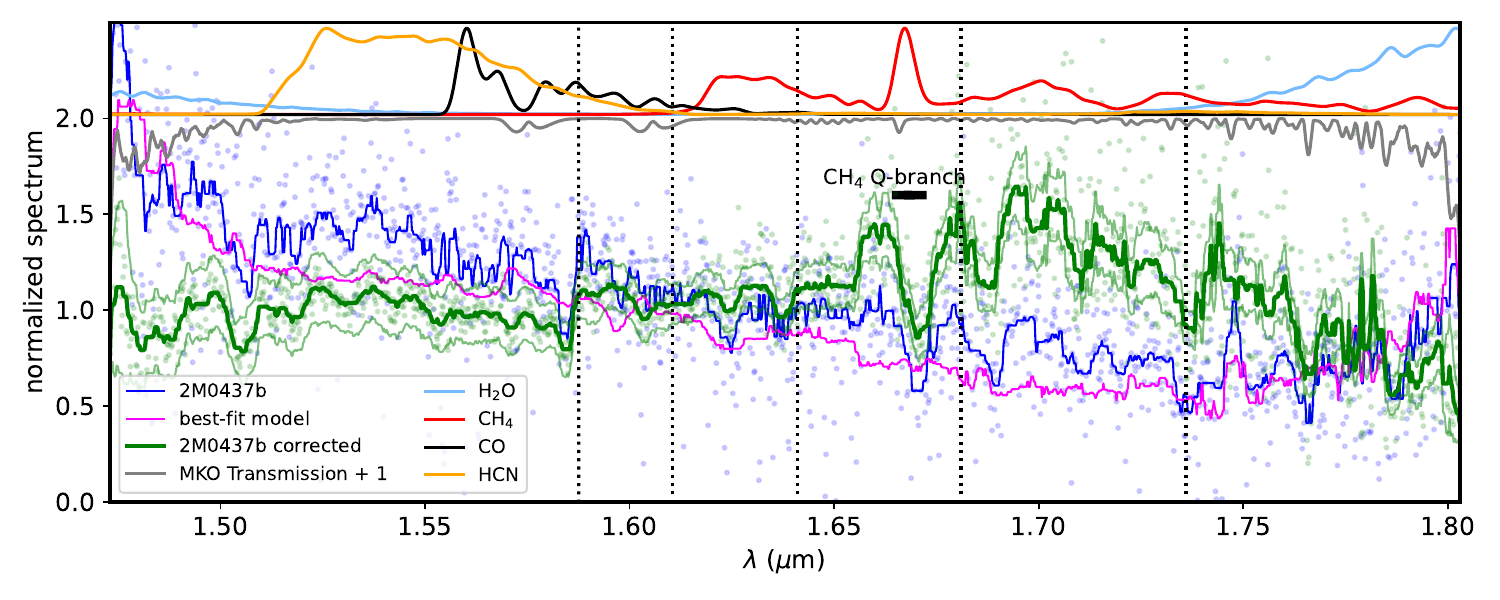}
 \includegraphics[width=\textwidth]{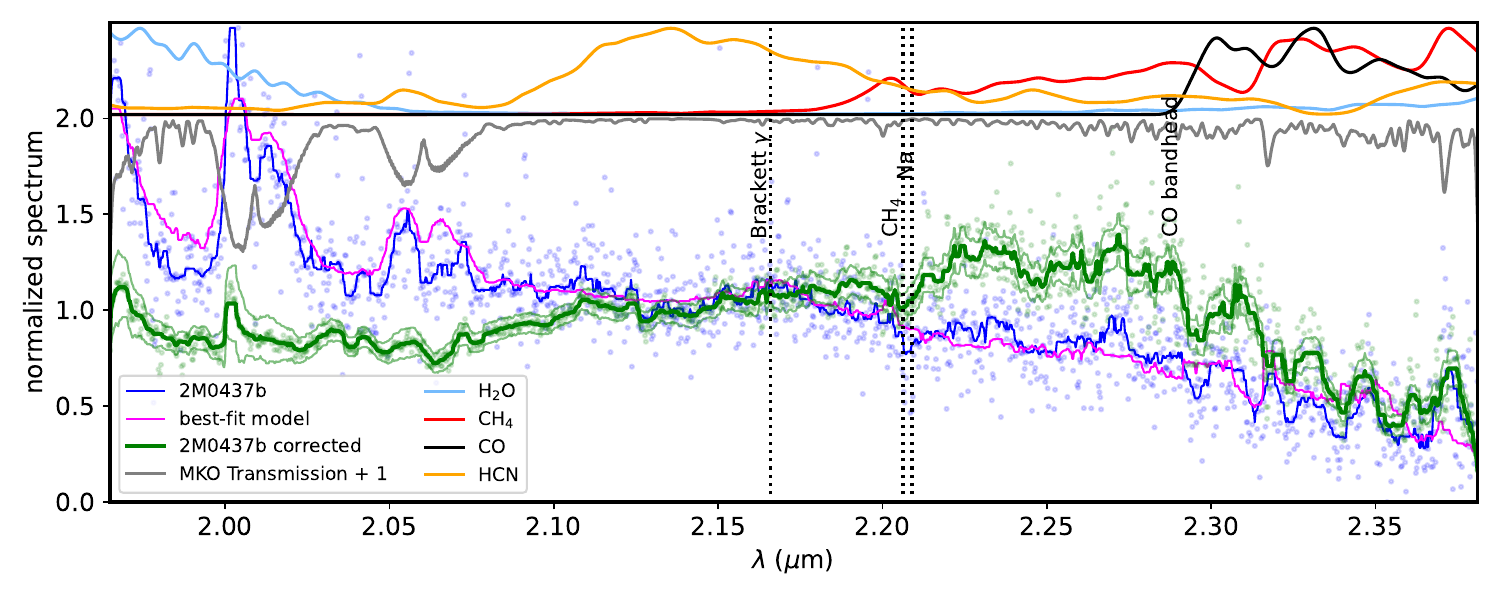}
    \caption{$H$-band (top) and $K$-band (bottom) spectra of 2M0437b.  The dark blue points are the reduced spectrum produced by the OSIRIS pipeline, the dark blue line is the 21-point running median, the magenta curve is a model of the spectrum to further correct for telluric absorption using a spectrum of an A0 star, and the green points and line are the corrected spectrum and its 21-point running median.  The grey line is a model transmission spectrum for the atmosphere over Maunakea for an airmass of 1.5 and water vapor column of 1.6 mm \citep[courtsey Gemini Orbservatory,][]{Lord1992}.  The absorption functions of key molecules (H$_2$O, CO, CH$_4$, HCN) computed using the {\tt HITRAN} database \citep{Mikhailenko2005,Gordon2022} and normalized by the maximum for that molecule in each bandpass are plotted at the top (blue, black, red, and orange lines, respectively.  The vertical dotted lines mark H Brackett lines and the Na\,I doublet, and spectral features of molecules of interest are labeled.}
    \label{fig:HKspec}
\end{figure*}

Figure \ref{fig:ircs} shows the extracted IRCS $YJ$ spectrum of 2M0437b, the background scattered light from the host star, and the A0-type star used for telluric correction.  The spectrum is of very low signal-to-noise and appears featureless.  Figure \ref{fig:HKspec} shows the $H$- and $K$-band spectra produced by the OSIRIS DRP (blue points), a 21-element median filtered version (dark blue line), and a spectrum corrected further for tellurics using the A0 star spectrum (green points and line).  The light green lines are the counting-noise uncertainties.  The overall shape of the spectrum is affected by imperfect correction for telluric absorption in the wings of each bandpass, especially on the blue side of the Kbb bandpass. 

For the purpose of feature identification, a representative telluric transmission spectrum (grey line) and absorption functions for key gases (H$_2$O, CO, CH$_4$, and HCN represented as light blue, black, red, and orange lines in the upper panels of Fig. \ref{fig:HKspec}) computed using {\tt HITRAN on the Web} \citep{Mikhailenko2005} and the 2020 version of the {\tt HITRAN} database \cite{Gordon2022} are shown.  The {\tt HITRAN} simulations were performed for 1200K (the companion's approximate \teff) and 0.2 bar \citep[the typical pressure at the radiative-convective boundary in a planetary atmosphere,]{}{Robinson2012} and the resolution was degraded by a Gaussian to emulate the median filtering performed on the data.      

The Kbb spectrum contains unambiguous evidence for atmospheric CO in the form of the bandhead of the overtone absorption at 2.3 \micron.  There is also evidence for CH$_4$, namely the Q-branch feature near 1.67 \micron\ and $\nu_2+\nu_3$ band at 2.2 \micron, although the latter also coincides with the the Na\,I doublet (Fig. \ref{fig:HKspec}).   These features appear in the individual spectra that were extracted from image mosaics and co-added, and they also appear in spectra constructed using different pipeline settings, although the strength of the features varies.   There is also tentative evidence for H$_2$O in the form of the downturn (absorption) at the red end of the $H$-band spectrum and blue end of the $K$-band spectrum.    

We do not see emission in H lines (i.e. Paschen $\beta$ or Brackett $\gamma$) as an indicator of ongoing accretion from a circumplanetary disk (Figs. \ref{fig:ircs} and \ref{fig:HKspec}).  Although there is a rise in the uncorrected spectrum of 2M0437b at 2.17$\mu$m, this is likely an artefact of the \emph{absorption} feature in the spectrum of the A0 star and is removed by our correction (Fig. \ref{fig:HKspec}).  

\section{Comparison with Other Objects and Models}
\label{sec:models}

\begin{figure*}
	\includegraphics[width=\textwidth]{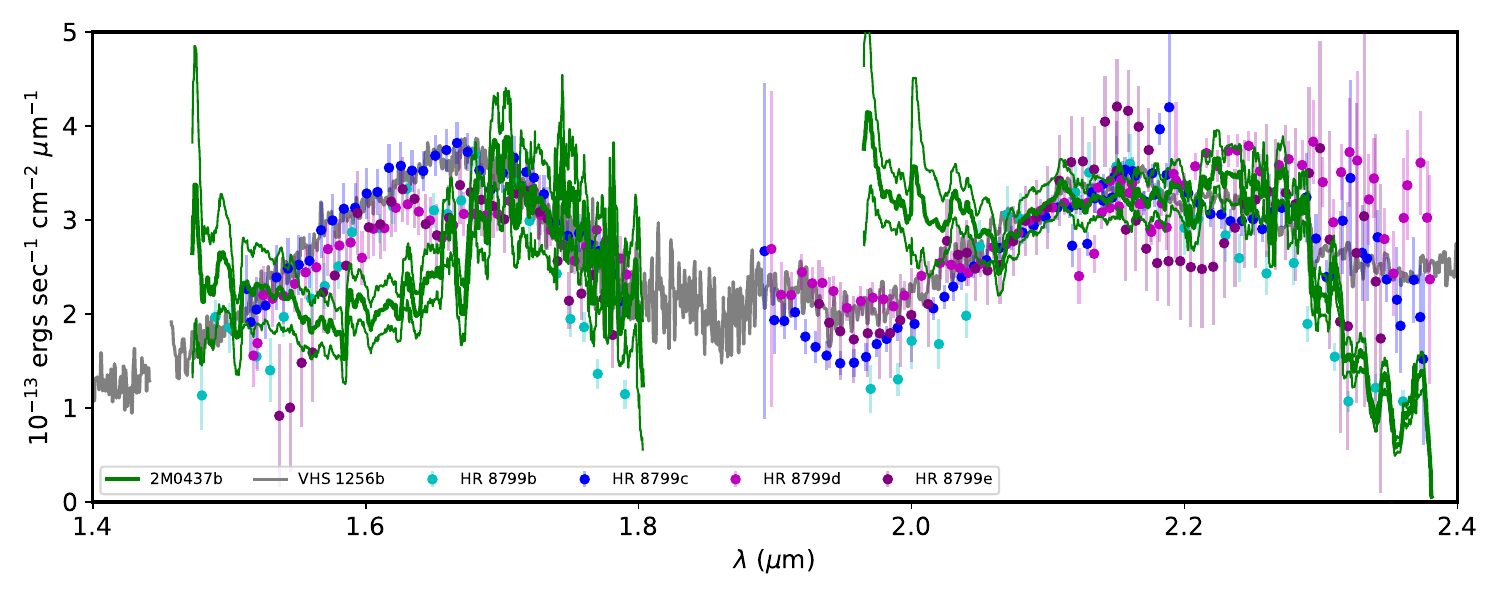}
    \caption{Comparison of the derived spectrum of 2M0437b (solid green line with light green lines denoting $\pm1\sigma$) with those of the bcde planets of HR 8799  \citep{Barman2015,Greenbaum2018} and VHS 1256b \citep[with \jwst-NIRSpec,][]{Miles2023}.  The comparison spectra have been re-normalized to agree with that of 2M0437b over the interval 2.05-2.15 \micron.}  
    \label{fig:compare}
\end{figure*}

We compared the OSIRIS $HK$ spectrum of 2M0437b to those of other planetary-mass objects that have similar $K$-band luminosities and $H-K$ colors \citep{Gaidos2022b}, namely the four known planets (bcde) of the HR 8799 system.  The \teff\ derived for all four planets is $\approx$1200K or slightly hotter in the case of c and e \citep{Greenbaum2018,Ruffio2021}, but these estimates are model-dependent.  \citet{Marois2008} and \citet{Currie2011} estimated their masses at 5, 7, 7, and 8 \mjup, respectively.  However, \citet{Sepulveda2022} revised the age of the HR 8799 system downwards from 30-60 to 10-20 Myr implying, all else being equal, lower planet masses.  CO and \water\ have been detected in spectra of HR 8799bcd but the presence of \methane\ is controversial \citep{Roche2018,Ruffio2021}. 
$HK$ spectra obtained with OSIRIS \citep[for HR 8799b][]{Barman2015} and GPI \citep[for cde][]{Greenbaum2018} have been re-normalized such that the median value in the $\lambda=2.05-2.15$ \micron\ window are equated and plotted along with 2M0437b in Fig. \ref{fig:compare}.  The overall agreement, particularly with HR 8799b, is notable.  The deviation in the blue side of the $K$-band is due to systematics from the telluric correction.  The pronounced deficit in the blue half of the $H$-band spectrum of 2M0437b is not seen in other spectra (or models) and its origin is discussed below.   Figure \ref{fig:compare} also compares 2M0437b with the re-normalized \jwst\ NIRSpec spectrum of VHS 1256-1257b from \citet{Miles2023}, which has a \teff\ of about 1100K, depending on method and models used, an age of $\sim$140 Myr, and a mass near or above the deuterium-burning limit \citep[13\mjup,][]{Dupuy2023,Miles2023}.     There are notable differences in the enhanced (reduced) relative emission in the \water-regions (windows), suggesting that VHS 1256b is much dustier than younger objects.  Another object with a similar luminosity/color is HIP 65426b (not shown) a $\sim$8 \mjup\-mass object with \teff$\approx$1500 orbiting a star in the $\approx$10-20 Myr-old Centaurus-Crux association \citep{Chauvin2017,Cheetham2019,Petrus2021}.

\begin{figure*}
	\includegraphics[width=\textwidth]{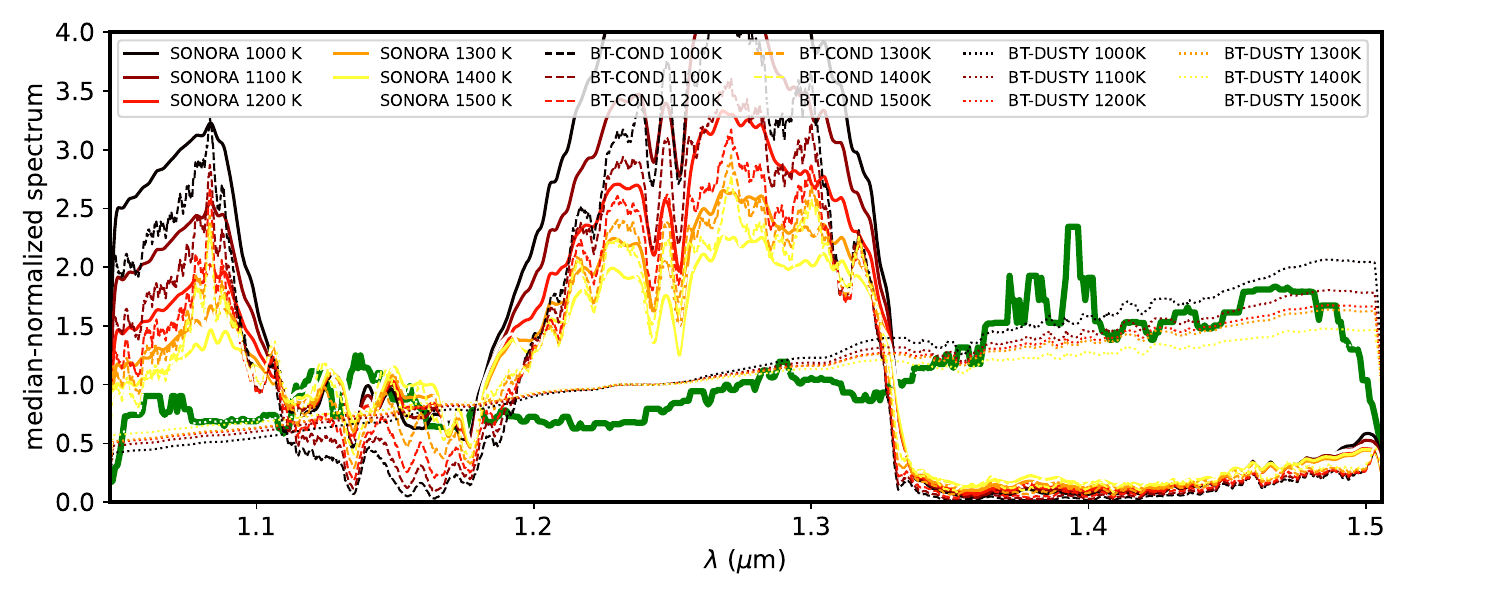}
    \caption{Comparison of IRCS $YJ$ spectrum (green curve) of 2M0437b with model spectra that ignore dust formation \citep[SONOROA BOBCAT,][]{Marley2021}, neglect the effects of dust on radiative transfer \citep[BT-COND][]{Allard2011}, and a model with a maximal dust effects \citep[BT-DUSTY,][]{Allard2011}.}  
    \label{fig:models_ircs}
\end{figure*}

\begin{figure*}
    \includegraphics[width=\textwidth]{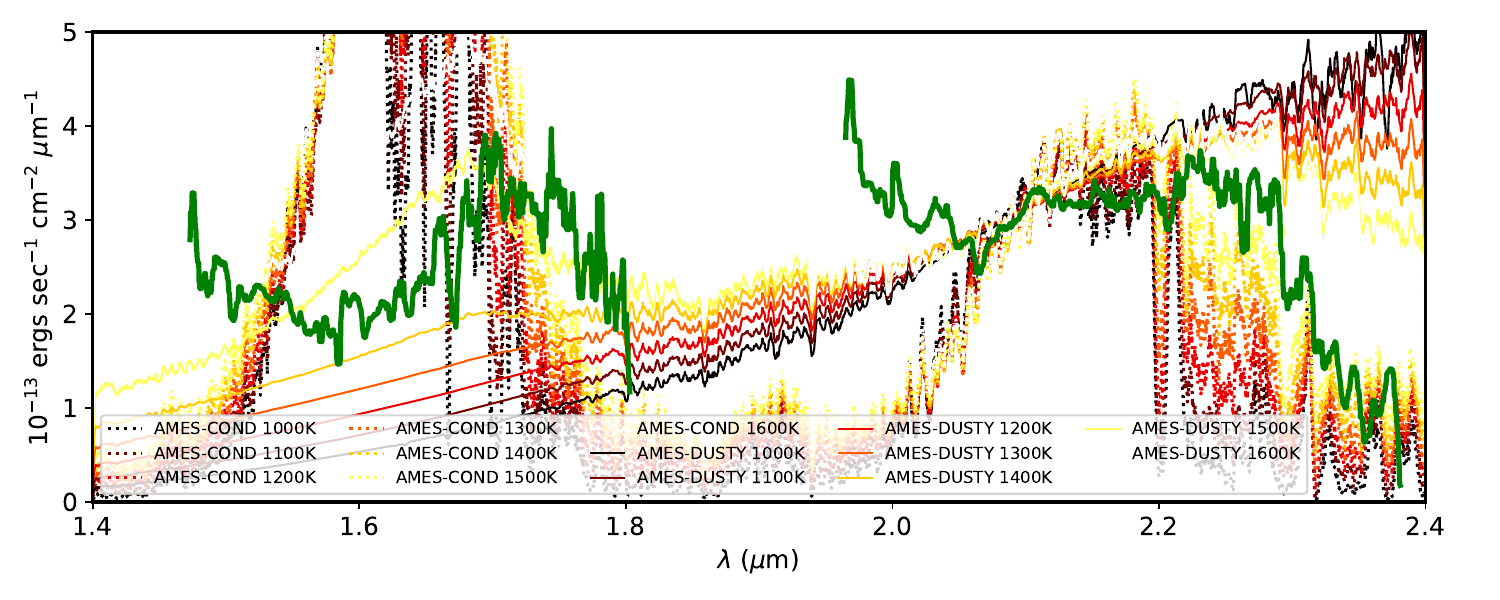} 
    \includegraphics[width=\textwidth]{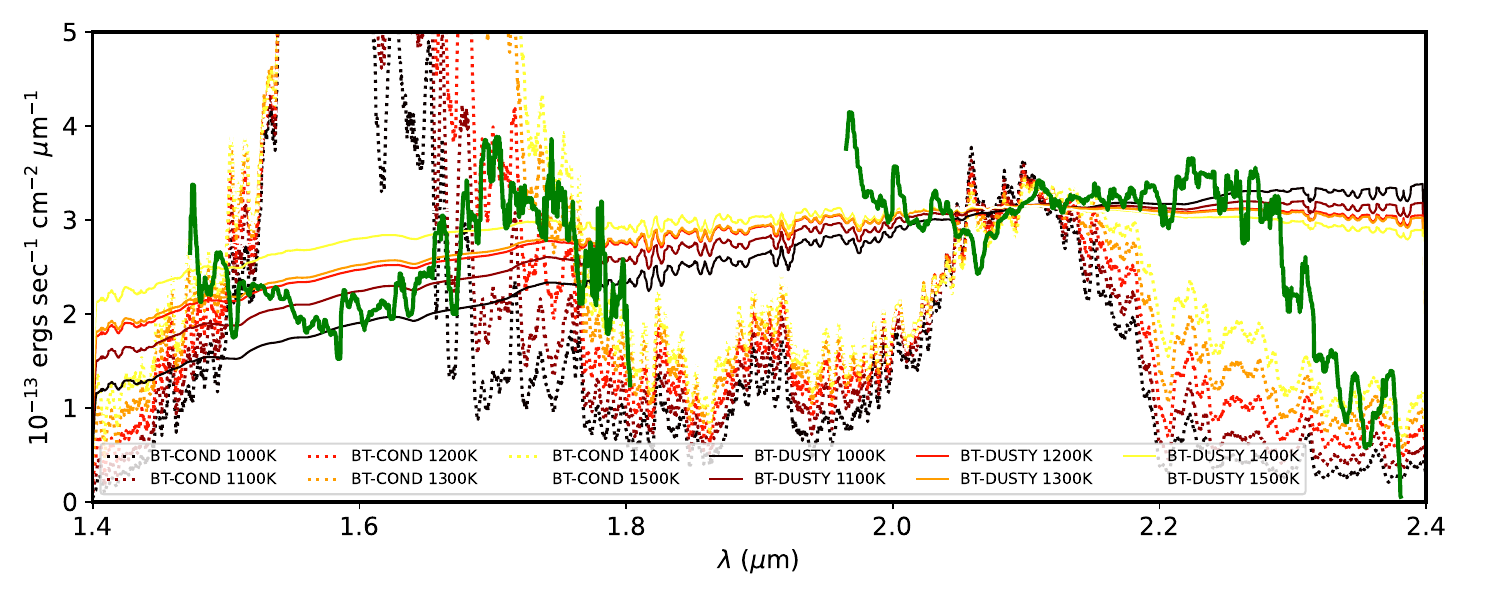} 
    \includegraphics[width=\textwidth]{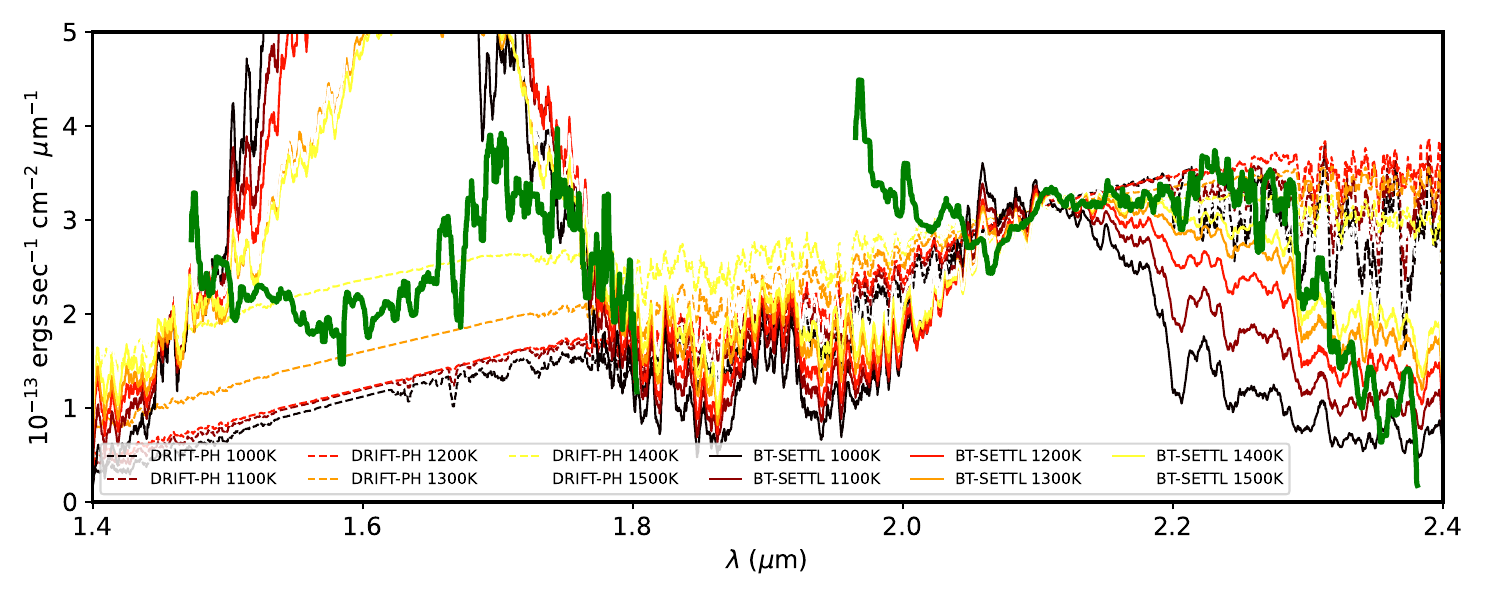}
    \caption{Comparison of OSIRIS $H$- and $K$-band spectra of 2M0437b with predictions of different sets of atmosphere models, all assuming solar metallicity and solar-like C/O.  In all cases, models are normalized to the median of the flux density in the middle of the $K$-band window ($\lambda\lambda$ = 2.05--2.15).  \textbf{Top:} AMES-COND (complete dust settling) vs. AMES-DUSTY (no dust settling) models, both sets with $log g = 3.5$ \citep{Chabrier2000}.  \textbf{Middle:} BT-COND (complete dust settling), BT-DUSTY (no dust settling), both with $log g = 4.5$, the lowest gravity cases computed.  Bottom: DRIFT-PHOENIX \citep{Woitke2004} and BT-SETTL \citep{Allard2012} models with models of dust settling based on atmospheric convection simulations, with $\log g$ of 3.5 and 4.5, respectively.}  
    \label{fig:models}
\end{figure*}

We compared our IRCS and OSIRIS spectra with published model spectra to infer properties of the object (Figs. \ref{fig:models_ircs} and \ref{fig:models}).  We adopted models with a solar metallicity, the value estimate for the Taurus star-forming region \citep{DOrazi2011}.  For the estimated mass ($\sim$4 \mjup) and age ($\sim$2-3 Myr) of 2M0437b \citep{Gaidos2022c}, models predict a gravity of $\log g \approx3.6$; we used $\log g = 3.5$ where available but otherwise adopted the lowest value from a suite.   

Two of the most important aspects of any model of emission from ultracool dwarf stars and luminous giant planets are its molecular line list and model for dust formation and settling (also termed ``rainout").  While the AMES series of models \citep{Chabrier2000} use the \citet{Partridge1997} line lists \citep{Chabrier2000}, the BT series of models are computed with PHOENIX stellar atmosphere models, which incorporate updated line lists, crucially for \water\ from \citet{Barber2006}.  Ongoing updates to line opacities have been incorporated in more recent models such as the cloud-free SONORA series \citep{Marley2021}.  These models typically adopt one of two extrema in treatments of condensates (dust/clouds).  In models like AMES-COND and BT-COND, settling of dust is assumed to be very efficient, effectively removing it from upper layers.  In the AMES-DUSTY and BT-DUSTY models, dust settling is ignored, and dust exists in equilibrium with local gas chemistry \citep{Allard2012}.  

More recent models have included more realistic cloud models.  The DRIFT-PHOENIX models \citep[][and references therein]{Woitke2004} include a detailed models of dust growth and settling incorporating a convective turnover time based on 3-d convection simulations \citep{Ludwig2002}.  The BT-SETTL models relates dust settling to simulated radiative-convective dynamics \citep{Freytag2010}, and includes detailed grain growth and optical properties \citep{Allard2012}.  There are a variety of approaches and trade-offs to preforming the chemical, radiative opacity, and structural calculations \citep[e.g.,][]{Marley2021}, we restrict our comparison to the AMES, BT, and SONORA models.

The near-infrared spectra predicted by cloud-free models with \teff\ $\lesssim$ 2000K are dominated by the absorption ro-vibrational bands of a small number of molecular species, particularly \water, leaving broad ``windows" through which most flux is radiated.  Clouds markedly increase the opacity in these windows, thereby reddening and smoothing the object's spectral appearance; the effect increases with increasing cloud height \citep[e.g., Fig. 2 in ][]{Brock2021}.   The featureless IRCS $YJ$ spectrum and subdued OSIRIS $H$-band spectrum of 2M0437b are clearly inconsistent with clear atmosphere models or those with efficient dust settling of dust (Figs. \ref{fig:models_ircs}  and \ref{fig:models}).   Indeed, \emph{none} of the dusty or dust-free models are in good agreement with the observed spectra of 2M0437b.  $\chi^2$ values were computed for models relative to the spectra over the intervals $>1.67$\micron\ in the $H$-band and $>2.07$\micron\ in the $K$-band to avoid the regions of potential systematics and uncorrected tellurics (or un-modeled absorption).  The best-fit models have \teff\ of 1500K for AMES-COND, 1900K for AMES-DUSTY, 1800K for BT-COND, 2000K for BT-DUSTY, 1100K for BT-SETTL and 2000K for DRFT-PHOENIX.  The lowest overall $\chi^2$ values are for AMES-COND and BT-SETTL, suggesting a \teff\ of $\sim$1200K, but the poor fit of all models precludes a more systematic estimate of temperature.   

The most pronounced departure from the model predictions is the deficit of $H$-band emission blue-ward of 1.7 \micron to at least 1.5 \micron.  This is a well-studied transmission``window" in the atmospheres of ultracool dwarfs between region of \water-dominated absorption, where models of clear atmospheres predict significant emission (e.g., SONORA BOBCAT models in Fig. \ref{fig:models_ircs}.)  Neither CO nor CH$_4$ are plausible sources of opacity in this region (Fig. \ref{fig:HKspec}).  Possible, non-exclusive explanations are an error in spectral extraction or correction/calibration, clouds/dust; and/or an additional molecular absorber.  The spectra extracted from two independent (but sequential) sets of observations in $H$-band show the same flux deficit.   Spectra extracted with a different rectification matrix obtained one year \emph{after} the observations also shows the deficit.  It is possible that some of the deficit is a result of the simple linear model for telluric correction, but the flat shape of this correction cannot explain the downturn at $\lambda < 1.7$\micron.  Absorption by \methane could explain at least some of the deficit at 1.6-1.7\micron, but not at shorter wavelengths.

\section{Discussion and Summary}
\label{sec:discussion}

\begin{figure}
	\includegraphics[width=\columnwidth]{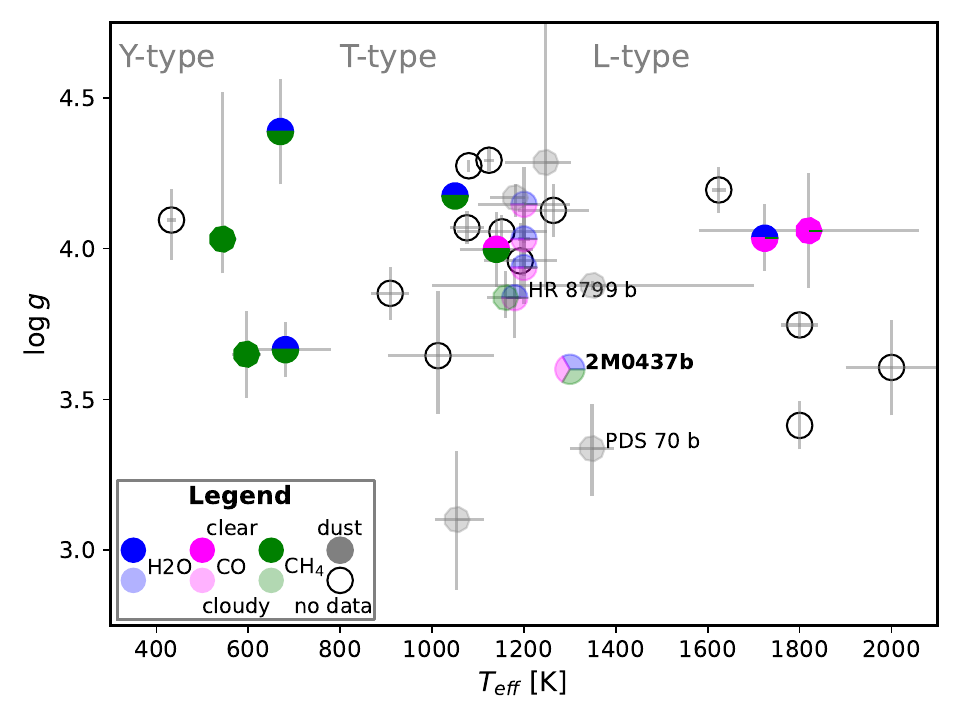}
    \caption{Gravity ($\log g$ cm~s$^{-2}$) vs. effective temperature \teff\ of 2M0437b vs. other confirmed free-floating or companion objects with estimated masses $<13 M_J$ and published radii and \teff.  2M0437b and its two most analogous counterparts are labeled.  Spectroscopic information on the atmosphere, if available, is represented by a pie chart, with colors indicating molecular constituents, light shading representing the presence of clouds/dust, and grey representing featureless spectra only indicating dust.  Empty circles represent objects with no published results. Grey points represent cases where spectra of the atmospheres are featureless and only dust is inferred, and empty circles represent no published results.  Planet parameters are from \citet{Greenbaum2018,Ruffio2021,Sepulveda2022} for HR 8799 bcde, \citet{Mueller2018,Madurowicz2023} for 51 Eri b, \citet{Desgrange2022} for HD 98056 b, \citet{Wang2021} for PDS 70 bc,
       \citet{Mesa2023} for AF Lep b, \citet{Schneider2023} for WISE J050626.96+073842.4, \citet{Skemer2016} for GJ 504 b, \citet{ZhangZ2021} for COCONUTS-2b,   \citet{Chilcote2017,Hoeijmakers2018,BrandtG2021} for $\beta$ Pic b, 
    \citet{Fontanive2020} for Oph 98 b, \citet{Naud2014} for GU Psc b, 
    \citet{Dupuy2013} for Ross 458 (AB) c, \citet{Vos2022} for 2MASS J0718-6415, J2117-2940, J0642+4101,  and J0642+4101, and WISE J2216+1952, \citet{Currie2014} for ROXs (AB) c, and \citet{Liu2013} and \citet{Miles2018} for PSO J318.5-22.}
    \label{fig:planets}
\end{figure}

$YJHK$ spectra of 2M0437b, the planetary-mass companion to a Taurus M dwarf, contain unambiguous evidence for atmospheric CO, possibly \water\ and \methane, and dust.  
Figure \ref{fig:planets} compares the ``surface" gravity $\log g$ and \teff\ of 2M0437b to those of other planetary-mass ($<$13\mjup\footnote{The choice of the D-burning limit of 13\mjup\ is a matter of convenience for selecting companions of similar mass and does not imply that higher-mass objects are not ``planets".}) companions or free-floating objects, with the colors/shading of the points keyed to spectroscopic information about the atmospheres.  2M0437b, like PDS 70 b and c, lies at the L- to T-type transition between hotter, dustier atmospheres and cooler, clearer atmospheres and is among the objects with the lowest gravity.

\water\ absorption bands are expected to dominate the overall spectral shape of an ultracool (\teff $<2600$K) with a solar-like metallicity,   However, these features are not apparent in our $YJ$ spectrum of 2M0437b and are attenuated in our $HK$ spectra, possibly due to micron-sized dust that scatters more effectively at shorter wavelengths.  A spectrum with weakened \water\ features and C-containing molecules could also be characteristic of gas giants that arise from an O-depleted disk in which condensation and inward migration of water ice (in the form of mm-sized pebbles or larger planetesimals) is efficient \citep{Schneider2021}, as suggested by high C/O in the gas of some disks \citep{Bosman2021}, and an organic-rich inner protoplanetary disk around an M dwarf \citep{Tabone2023}.  

The pronounced overtone bandhead of CO at 2.3 \micron\ is characteristic of L dwarfs (\teff $<$2200K) but is less distinct among cooler T dwarfs \citep[e.g.,][]{Geballe2002}.  The L-T transition in ultracool dwarfs, at 1200-1400K, is accompanied by the emergence of strong \methane\ lines explainable by the disappearance or ``sinking" of dust to deeper layers in the atmosphere \citep{Marley2015}.  Methane features typically become strong in T dwarf atmospheres but can be present in late L-type atmospheres \citep{Geballe2002} due to dust in deep layers.  Absorption by both CO and CH$_4$, and scattering by dust points to a \teff{} near or within this transition, consistent with the luminosity-based estimate of 1400-1500K by \citet{Gaidos2022b}.  The suppressed flux at wavelengths shorter than 1.6 \micron, if not an artefact, could be due to other absorbers including, speculatively, HCN \citep[orange lines in Fig. \ref{fig:HKspec},][]{MacDonald2017}, or PAHs \citep{ChenT2019}.  

Since the system's age has been estimated at $\lesssim5$ Myr \citep{Gaidos2022b} and giant planet formation (at least by the core-first scenario) might take up to a few Myr \citep{Helled2014}, it is possible that 2M0437b is \emph{exceptionally} young.  Because of its extreme youth, the spectrum of 2M0437b might differ substantially from its older counterparts with the same mass or \teff.  Firstly, the object would be in the early stages of contraction, have a larger radius and lower gravity, and less efficient dust settling -- the persistence of dust clouds is thought to impart the characteristic redness of such objects \citep{Liu2016}.  Second, it is possible that the planet is still accreting, although we do not see evidence for this in the form of H line emission in the spectrum.  Third, the object could be still embedded in a circumstellar disk or other envelope which, due to planet migration or chemical processing, has a distinct chemical composition and dust population.  There is no evidence for a residual dust/gas disk around 2M0437, although quasi-periodic dimming events observed by the \ktwo\ mission were probably caused by close-in occulting dust clouds \citep{Gaidos2022b}.

\citet{Desgrange2022} found that that the red spectral energy distribution of the substellar companion to the $\sim$13 Myr star HD 95086 could be explained either if ``b" is a hotter object with extremely high circumplanetary extinction (following a standard ISM extinction law) or a significantly cooler object with moderate extinction.  Low-resolution IR spectra of the two super-Jupiters embedded in the disk around the $\sim$5 Myr-old T Tauri star PDS 70 are only consistent with existing models if there is \emph{additional} extinction by dust (above the atmosphere) of at least several magnitudes \citep{Wang2021}.  A $K$-band spectrum of the inner of the two known giant planets around the $\sim$25 Myr-old $\beta$ Pictoris appears \emph{completely} featureless and also indicates very high extinction along the line of sight \citep{Cugno2021}.  

2M0437b is a challenging but appropriate target for \jwst\ observations to obtain moderate resolution ($\gtrsim$1000) spectra with the NIRSpec and/or MIRI integral field spectrographs, covering 1-5.3 \micron\ at 0".1 spatial resolution, and 5-28 \micron\ at 0".3 pixel scale, respectively.  A \jwst\ NIRSpec+MIRI spectrum of VHS 1256–1257b, with an analogous L-to-T transition spectral type but more massive and older ($\sim$140 Myr) than 2M0437b, revealed molecular features in detail as well as lines of atomic Na and K \citep{Miles2023}.  A circumplanetary disk could also be detected via excess emission in the mid-IR.

\section*{Acknowledgements}

We thank Sherry Yeh, Randy Campbell, and Jim Lyke of the W. M. Keck Observatory for their multi-faceted assistance with OSIRIS observations and data reduction, and Mike Liu and Paul Molli\`{e}re for useful suggestions.  The data presented herein were obtained at the W. M. Keck Observatory, which is operated as a scientific partnership among the California Institute of Technology, the University of California and the National Aeronautics and Space Administration. The Observatory was made possible by the generous financial support of the W. M. Keck Foundation.  EG was supported by NASA Award 80NSSC20K0957 (Exoplanets Research Program).  T.H.\ was supported by JSPS KAKENHI Grant Numbers JP19K14783 and JP21H00035.  This research is also based in part on data collected at the Subaru Telescope, which is operated by the National Astronomical Observatory of Japan. This research made use of {\tt Astropy}\footnote{http://www.astropy.org} a community-developed core Python package for Astronomy \citep{Astropy2013,astropy:2018}.   

\section*{Data availability}  

The IRCS and OSIRIS data used in this work are publicly available on the SMOKA archive (https://smoka.nao.ac.jp/) and KOA archive (https://www2.keck.hawaii.edu/koa/public/koa.php), respectively.





\begin{thebibliography}{}
\makeatletter
\relax
\def\mn@urlcharsother{\let\do\@makeother \do\$\do\&\do\#\do\^\do\_\do\%\do\~}
\def\mn@doi{\begingroup\mn@urlcharsother \@ifnextchar [ {\mn@doi@}
  {\mn@doi@[]}}
\def\mn@doi@[#1]#2{\def\@tempa{#1}\ifx\@tempa\@empty \href
  {http://dx.doi.org/#2} {doi:#2}\else \href {http://dx.doi.org/#2} {#1}\fi
  \endgroup}
\def\mn@eprint#1#2{\mn@eprint@#1:#2::\@nil}
\def\mn@eprint@arXiv#1{\href {http://arxiv.org/abs/#1} {{\tt arXiv:#1}}}
\def\mn@eprint@dblp#1{\href {http://dblp.uni-trier.de/rec/bibtex/#1.xml}
  {dblp:#1}}
\def\mn@eprint@#1:#2:#3:#4\@nil{\def\@tempa {#1}\def\@tempb {#2}\def\@tempc
  {#3}\ifx \@tempc \@empty \let \@tempc \@tempb \let \@tempb \@tempa \fi \ifx
  \@tempb \@empty \def\@tempb {arXiv}\fi \@ifundefined
  {mn@eprint@\@tempb}{\@tempb:\@tempc}{\expandafter \expandafter \csname
  mn@eprint@\@tempb\endcsname \expandafter{\@tempc}}}

\bibitem[\protect\citeauthoryear{{Allard}, {Homeier}  \& {Freytag}}{{Allard}
  et~al.}{2011}]{Allard2011}
{Allard} F.,  {Homeier} D.,   {Freytag} B.,  2011, in {Johns-Krull} C.,
  {Browning} M.~K.,   {West} A.~A.,  eds,  Astronomical Society of the Pacific
  Conference Series Vol. 448, 16th Cambridge Workshop on Cool Stars, Stellar
  Systems, and the Sun. p.~91 (\mn@eprint {arXiv} {1011.5405})

\bibitem[\protect\citeauthoryear{{Allard}, {Homeier}  \& {Freytag}}{{Allard}
  et~al.}{2012}]{Allard2012}
{Allard} F.,  {Homeier} D.,   {Freytag} B.,  2012, \mn@doi [Philosophical
  Transactions of the Royal Society of London Series A]
  {10.1098/rsta.2011.0269}, \href
  {https://ui.adsabs.harvard.edu/abs/2012RSPTA.370.2765A} {370, 2765}

\bibitem[\protect\citeauthoryear{{Astropy Collaboration} et~al.,}{{Astropy
  Collaboration} et~al.}{2013}]{Astropy2013}
{Astropy Collaboration} et~al., 2013, \mn@doi [\aap]
  {10.1051/0004-6361/201322068}, \href
  {http://adsabs.harvard.edu/abs/2013A%26A...558A..33A} {558, A33}

\bibitem[\protect\citeauthoryear{{Barber}, {Tennyson}, {Harris}  \&
  {Tolchenov}}{{Barber} et~al.}{2006}]{Barber2006}
{Barber} R.~J.,  {Tennyson} J.,  {Harris} G.~J.,   {Tolchenov} R.~N.,  2006,
  \mn@doi [\mnras] {10.1111/j.1365-2966.2006.10184.x}, \href
  {https://ui.adsabs.harvard.edu/abs/2006MNRAS.368.1087B} {368, 1087}

\bibitem[\protect\citeauthoryear{{Barman}, {Konopacky}, {Macintosh}  \&
  {Marois}}{{Barman} et~al.}{2015}]{Barman2015}
{Barman} T.~S.,  {Konopacky} Q.~M.,  {Macintosh} B.,   {Marois} C.,  2015,
  \mn@doi [\apj] {10.1088/0004-637X/804/1/61}, \href
  {https://ui.adsabs.harvard.edu/abs/2015ApJ...804...61B} {804, 61}

\bibitem[\protect\citeauthoryear{{Bosman} et~al.,}{{Bosman}
  et~al.}{2021}]{Bosman2021}
{Bosman} A.~D.,  et~al., 2021, \mn@doi [\apjs] {10.3847/1538-4365/ac1435},
  \href {https://ui.adsabs.harvard.edu/abs/2021ApJS..257....7B} {257, 7}

\bibitem[\protect\citeauthoryear{{Bowler}, {Zhou}, {Morley}, {Kataria},
  {Bryan}, {Benneke}  \& {Batygin}}{{Bowler} et~al.}{2020}]{Bowler2020b}
{Bowler} B.~P.,  {Zhou} Y.,  {Morley} C.~V.,  {Kataria} T.,  {Bryan} M.~L.,
  {Benneke} B.,   {Batygin} K.,  2020, \mn@doi [\apjl]
  {10.3847/2041-8213/ab8197}, \href
  {https://ui.adsabs.harvard.edu/abs/2020ApJ...893L..30B} {893, L30}

\bibitem[\protect\citeauthoryear{{Brandt}, {Brandt}, {Dupuy}, {Li}  \&
  {Michalik}}{{Brandt} et~al.}{2021}]{BrandtG2021}
{Brandt} G.~M.,  {Brandt} T.~D.,  {Dupuy} T.~J.,  {Li} Y.,   {Michalik} D.,
  2021, \mn@doi [\aj] {10.3847/1538-3881/abdc2e}, \href
  {https://ui.adsabs.harvard.edu/abs/2021AJ....161..179B} {161, 179}

\bibitem[\protect\citeauthoryear{{Brock}, {Barman}, {Konopacky}  \&
  {Stone}}{{Brock} et~al.}{2021}]{Brock2021}
{Brock} L.,  {Barman} T.,  {Konopacky} Q.~M.,   {Stone} J.~M.,  2021, \mn@doi
  [\apj] {10.3847/1538-4357/abfc46}, \href
  {https://ui.adsabs.harvard.edu/abs/2021ApJ...914..124B} {914, 124}

\bibitem[\protect\citeauthoryear{{Chabrier}, {Baraffe}, {Allard}  \&
  {Hauschildt}}{{Chabrier} et~al.}{2000}]{Chabrier2000}
{Chabrier} G.,  {Baraffe} I.,  {Allard} F.,   {Hauschildt} P.,  2000, \mn@doi
  [\apj] {10.1086/309513}, \href
  {https://ui.adsabs.harvard.edu/abs/2000ApJ...542..464C} {542, 464}

\bibitem[\protect\citeauthoryear{{Chauvin} et~al.,}{{Chauvin}
  et~al.}{2017}]{Chauvin2017}
{Chauvin} G.,  et~al., 2017, \mn@doi [\aap] {10.1051/0004-6361/201731152},
  \href {https://ui.adsabs.harvard.edu/abs/2017A&A...605L...9C} {605, L9}

\bibitem[\protect\citeauthoryear{{Cheetham} et~al.,}{{Cheetham}
  et~al.}{2019}]{Cheetham2019}
{Cheetham} A.~C.,  et~al., 2019, \mn@doi [\aap] {10.1051/0004-6361/201834112},
  \href {https://ui.adsabs.harvard.edu/abs/2019A&A...622A..80C} {622, A80}

\bibitem[\protect\citeauthoryear{{Chen} \& {Szul{\'a}gyi}}{{Chen} \&
  {Szul{\'a}gyi}}{2022}]{ChenX2022}
{Chen} X.,  {Szul{\'a}gyi} J.,  2022, \mn@doi [\mnras]
  {10.1093/mnras/stac1976}, \href
  {https://ui.adsabs.harvard.edu/abs/2022MNRAS.516..506C} {516, 506}

\bibitem[\protect\citeauthoryear{{Chen}, {Luo}  \& {Li}}{{Chen}
  et~al.}{2019}]{ChenT2019}
{Chen} T.,  {Luo} Y.,   {Li} A.,  2019, \mn@doi [\aap]
  {10.1051/0004-6361/201936310}, \href
  {https://ui.adsabs.harvard.edu/abs/2019A&A...632A..71C} {632, A71}

\bibitem[\protect\citeauthoryear{{Chilcote} et~al.,}{{Chilcote}
  et~al.}{2017}]{Chilcote2017}
{Chilcote} J.,  et~al., 2017, \mn@doi [\aj] {10.3847/1538-3881/aa63e9}, \href
  {https://ui.adsabs.harvard.edu/abs/2017AJ....153..182C} {153, 182}

\bibitem[\protect\citeauthoryear{{Cridland}, {Eistrup}  \& {van
  Dishoeck}}{{Cridland} et~al.}{2019}]{Cridland2019}
{Cridland} A.~J.,  {Eistrup} C.,   {van Dishoeck} E.~F.,  2019, \mn@doi [\aap]
  {10.1051/0004-6361/201834378}, \href
  {https://ui.adsabs.harvard.edu/abs/2019A&A...627A.127C} {627, A127}

\bibitem[\protect\citeauthoryear{{Cugno} et~al.,}{{Cugno}
  et~al.}{2021}]{Cugno2021}
{Cugno} G.,  et~al., 2021, \mn@doi [\aap] {10.1051/0004-6361/202140632}, \href
  {https://ui.adsabs.harvard.edu/abs/2021A&A...653A..12C} {653, A12}

\bibitem[\protect\citeauthoryear{{Currie} et~al.,}{{Currie}
  et~al.}{2011}]{Currie2011}
{Currie} T.,  et~al., 2011, \mn@doi [\apj] {10.1088/0004-637X/729/2/128}, \href
  {https://ui.adsabs.harvard.edu/abs/2011ApJ...729..128C} {729, 128}

\bibitem[\protect\citeauthoryear{{Currie}, {Daemgen}, {Debes}, {Lafreniere},
  {Itoh}, {Jayawardhana}, {Ratzka}  \& {Correia}}{{Currie}
  et~al.}{2014}]{Currie2014}
{Currie} T.,  {Daemgen} S.,  {Debes} J.,  {Lafreniere} D.,  {Itoh} Y.,
  {Jayawardhana} R.,  {Ratzka} T.,   {Correia} S.,  2014, \mn@doi [\apjl]
  {10.1088/2041-8205/780/2/L30}, \href
  {https://ui.adsabs.harvard.edu/abs/2014ApJ...780L..30C} {780, L30}

\bibitem[\protect\citeauthoryear{{Currie}, {Biller}, {Lagrange}, {Marois},
  {Guyon}, {Nielsen}, {Bonnefoy}  \& {De Rosa}}{{Currie}
  et~al.}{2022}]{Currie2022}
{Currie} T.,  {Biller} B.,  {Lagrange} A.-M.,  {Marois} C.,  {Guyon} O.,
  {Nielsen} E.,  {Bonnefoy} M.,   {De Rosa} R.,  2022, \mn@doi [arXiv e-prints]
  {10.48550/arXiv.2205.05696}, \href
  {https://ui.adsabs.harvard.edu/abs/2022arXiv220505696C} {p. arXiv:2205.05696}

\bibitem[\protect\citeauthoryear{{D'Orazi}, {Biazzo}  \& {Randich}}{{D'Orazi}
  et~al.}{2011}]{DOrazi2011}
{D'Orazi} V.,  {Biazzo} K.,   {Randich} S.,  2011, \mn@doi [\aap]
  {10.1051/0004-6361/201015616}, \href
  {https://ui.adsabs.harvard.edu/abs/2011A&A...526A.103D} {526, A103}

\bibitem[\protect\citeauthoryear{{Desgrange} et~al.,}{{Desgrange}
  et~al.}{2022}]{Desgrange2022}
{Desgrange} C.,  et~al., 2022, \mn@doi [\aap] {10.1051/0004-6361/202243097},
  \href {https://ui.adsabs.harvard.edu/abs/2022A&A...664A.139D} {664, A139}

\bibitem[\protect\citeauthoryear{{Dupuy} \& {Kraus}}{{Dupuy} \&
  {Kraus}}{2013}]{Dupuy2013}
{Dupuy} T.~J.,  {Kraus} A.~L.,  2013, \mn@doi [Science]
  {10.1126/science.1241917}, \href
  {https://ui.adsabs.harvard.edu/abs/2013Sci...341.1492D} {341, 1492}

\bibitem[\protect\citeauthoryear{{Dupuy}, {Liu}, {Evans}, {Best}, {Pearce},
  {Sanghi}, {Phillips}  \& {Bardalez Gagliuffi}}{{Dupuy}
  et~al.}{2023}]{Dupuy2023}
{Dupuy} T.~J.,  {Liu} M.~C.,  {Evans} E.~L.,  {Best} W. M.~J.,  {Pearce} L.~A.,
   {Sanghi} A.,  {Phillips} M.~W.,   {Bardalez Gagliuffi} D.~C.,  2023, \mn@doi
  [\mnras] {10.1093/mnras/stac3557}, \href
  {https://ui.adsabs.harvard.edu/abs/2023MNRAS.519.1688D} {519, 1688}

\bibitem[\protect\citeauthoryear{{Fontanive} et~al.,}{{Fontanive}
  et~al.}{2020}]{Fontanive2020}
{Fontanive} C.,  et~al., 2020, \mn@doi [\apjl] {10.3847/2041-8213/abcaf8},
  \href {https://ui.adsabs.harvard.edu/abs/2020ApJ...905L..14F} {905, L14}

\bibitem[\protect\citeauthoryear{{Freytag}, {Allard}, {Ludwig}, {Homeier}  \&
  {Steffen}}{{Freytag} et~al.}{2010}]{Freytag2010}
{Freytag} B.,  {Allard} F.,  {Ludwig} H.~G.,  {Homeier} D.,   {Steffen} M.,
  2010, \mn@doi [\aap] {10.1051/0004-6361/200913354}, \href
  {https://ui.adsabs.harvard.edu/abs/2010A&A...513A..19F} {513, A19}

\bibitem[\protect\citeauthoryear{{Gaidos} et~al.,}{{Gaidos}
  et~al.}{2022a}]{Gaidos2022b}
{Gaidos} E.,  et~al., 2022a, \mn@doi [\mnras] {10.1093/mnras/stab3069}, \href
  {https://ui.adsabs.harvard.edu/abs/2022MNRAS.512..583G} {512, 583}

\bibitem[\protect\citeauthoryear{{Gaidos} et~al.,}{{Gaidos}
  et~al.}{2022b}]{Gaidos2022c}
{Gaidos} E.,  et~al., 2022b, \mn@doi [\mnras] {10.1093/mnras/stac1433}, \href
  {https://ui.adsabs.harvard.edu/abs/2022MNRAS.514.1386G} {514, 1386}

\bibitem[\protect\citeauthoryear{{Geballe} et~al.,}{{Geballe}
  et~al.}{2002}]{Geballe2002}
{Geballe} T.~R.,  et~al., 2002, \mn@doi [\apj] {10.1086/324078}, \href
  {https://ui.adsabs.harvard.edu/abs/2002ApJ...564..466G} {564, 466}

\bibitem[\protect\citeauthoryear{{Gordon} et~al.,}{{Gordon}
  et~al.}{2022}]{Gordon2022}
{Gordon} I.~E.,  et~al., 2022, \mn@doi [\jqsrt] {10.1016/j.jqsrt.2021.107949},
  \href {https://ui.adsabs.harvard.edu/abs/2022JQSRT.27707949G} {277, 107949}

\bibitem[\protect\citeauthoryear{{Greenbaum} et~al.,}{{Greenbaum}
  et~al.}{2018}]{Greenbaum2018}
{Greenbaum} A.~Z.,  et~al., 2018, \mn@doi [\aj] {10.3847/1538-3881/aabcb8},
  \href {https://ui.adsabs.harvard.edu/abs/2018AJ....155..226G} {155, 226}

\bibitem[\protect\citeauthoryear{{Hayano} et~al.,}{{Hayano}
  et~al.}{2010}]{Hayano2010}
{Hayano} Y.,  et~al., 2010, in {Ellerbroek} B.~L.,  {Hart} M.,  {Hubin} N.,
  {Wizinowich} P.~L.,  eds,  Society of Photo-Optical Instrumentation Engineers
  (SPIE) Conference Series Vol. 7736, Adaptive Optics Systems II. p. 77360N,
  \mn@doi{10.1117/12.857567}

\bibitem[\protect\citeauthoryear{{Helled} et~al.,}{{Helled}
  et~al.}{2014}]{Helled2014}
{Helled} R.,  et~al., 2014, in {Beuther} H.,  {Klessen} R.~S.,  {Dullemond}
  C.~P.,   {Henning} T.,  eds, Protostars and Planets VI. pp 643--665
  (\mn@eprint {arXiv} {1311.1142}),
  \mn@doi{10.2458/azu_uapress_9780816531240-ch028}

\bibitem[\protect\citeauthoryear{{Hoeijmakers}, {Schwarz}, {Snellen}, {de Kok},
  {Bonnefoy}, {Chauvin}, {Lagrange}  \& {Girard}}{{Hoeijmakers}
  et~al.}{2018}]{Hoeijmakers2018}
{Hoeijmakers} H.~J.,  {Schwarz} H.,  {Snellen} I.~A.~G.,  {de Kok} R.~J.,
  {Bonnefoy} M.,  {Chauvin} G.,  {Lagrange} A.~M.,   {Girard} J.~H.,  2018,
  \mn@doi [\aap] {10.1051/0004-6361/201832902}, \href
  {https://ui.adsabs.harvard.edu/abs/2018A&A...617A.144H} {617, A144}

\bibitem[\protect\citeauthoryear{{Kobayashi} et~al.,}{{Kobayashi}
  et~al.}{2000}]{Kobayashi2000}
{Kobayashi} N.,  et~al., 2000, in {Iye} M.,  {Moorwood} A.~F.,  eds,  Society
  of Photo-Optical Instrumentation Engineers (SPIE) Conference Series Vol.
  4008, Optical and IR Telescope Instrumentation and Detectors. pp 1056--1066,
  \mn@doi{10.1117/12.395423}

\bibitem[\protect\citeauthoryear{{Larkin} et~al.,}{{Larkin}
  et~al.}{2006}]{Larkin2006}
{Larkin} J.,  et~al., 2006, in {McLean} I.~S.,  {Iye} M.,  eds,  Society of
  Photo-Optical Instrumentation Engineers (SPIE) Conference Series Vol. 6269,
  Ground-based and Airborne Instrumentation for Astronomy. p. 62691A,
  \mn@doi{10.1117/12.672061}

\bibitem[\protect\citeauthoryear{{Liu} et~al.,}{{Liu} et~al.}{2013}]{Liu2013}
{Liu} M.~C.,  et~al., 2013, \mn@doi [\apjl] {10.1088/2041-8205/777/2/L20},
  \href {https://ui.adsabs.harvard.edu/abs/2013ApJ...777L..20L} {777, L20}

\bibitem[\protect\citeauthoryear{{Liu}, {Dupuy}  \& {Allers}}{{Liu}
  et~al.}{2016}]{Liu2016}
{Liu} M.~C.,  {Dupuy} T.~J.,   {Allers} K.~N.,  2016, \mn@doi [\apj]
  {10.3847/1538-4357/833/1/96}, \href
  {https://ui.adsabs.harvard.edu/abs/2016ApJ...833...96L} {833, 96}

\bibitem[\protect\citeauthoryear{{Lockhart} et~al.,}{{Lockhart}
  et~al.}{2019}]{Lockhart2019}
{Lockhart} K.~E.,  et~al., 2019, \mn@doi [\aj] {10.3847/1538-3881/aaf64e},
  \href {https://ui.adsabs.harvard.edu/abs/2019AJ....157...75L} {157, 75}

\bibitem[\protect\citeauthoryear{{Lord}}{{Lord}}{1992}]{Lord1992}
{Lord} S.~D.,  1992, {A new software tool for computing Earth's atmospheric
  transmission of near- and far-infrared radiation}, NASA Technical Memorandum
  103957

\bibitem[\protect\citeauthoryear{{Ludwig}, {Allard}  \& {Hauschildt}}{{Ludwig}
  et~al.}{2002}]{Ludwig2002}
{Ludwig} H.~G.,  {Allard} F.,   {Hauschildt} P.~H.,  2002, \mn@doi [\aap]
  {10.1051/0004-6361:20021153}, \href
  {https://ui.adsabs.harvard.edu/abs/2002A&A...395...99L} {395, 99}

\bibitem[\protect\citeauthoryear{{Lyke} et~al.,}{{Lyke}
  et~al.}{2017}]{Lyke2017}
{Lyke} J.,  et~al., 2017, {OSIRIS Toolbox: OH-Suppressing InfraRed Imaging
  Spectrograph pipeline}, Astrophysics Source Code Library, record
  ascl:1710.021 (\mn@eprint {ascl} {1710.021})

\bibitem[\protect\citeauthoryear{{MacDonald} \& {Madhusudhan}}{{MacDonald} \&
  {Madhusudhan}}{2017}]{MacDonald2017}
{MacDonald} R.~J.,  {Madhusudhan} N.,  2017, \mn@doi [\apjl]
  {10.3847/2041-8213/aa97d4}, \href
  {https://ui.adsabs.harvard.edu/abs/2017ApJ...850L..15M} {850, L15}

\bibitem[\protect\citeauthoryear{{Madurowicz}, {Mukherjee}, {Batalha},
  {Macintosh}, {Marley}  \& {Karalidi}}{{Madurowicz}
  et~al.}{2023}]{Madurowicz2023}
{Madurowicz} A.,  {Mukherjee} S.,  {Batalha} N.,  {Macintosh} B.,  {Marley} M.,
    {Karalidi} T.,  2023, \mn@doi [\aj] {10.3847/1538-3881/acca7a}, \href
  {https://ui.adsabs.harvard.edu/abs/2023AJ....165..238M} {165, 238}

\bibitem[\protect\citeauthoryear{{Marleau} \& {Cumming}}{{Marleau} \&
  {Cumming}}{2014}]{Marleau2014}
{Marleau} G.~D.,  {Cumming} A.,  2014, \mn@doi [\mnras]
  {10.1093/mnras/stt1967}, \href
  {https://ui.adsabs.harvard.edu/abs/2014MNRAS.437.1378M} {437, 1378}

\bibitem[\protect\citeauthoryear{{Marley} \& {Robinson}}{{Marley} \&
  {Robinson}}{2015}]{Marley2015}
{Marley} M.~S.,  {Robinson} T.~D.,  2015, \mn@doi [\araa]
  {10.1146/annurev-astro-082214-122522}, \href
  {https://ui.adsabs.harvard.edu/abs/2015ARA&A..53..279M} {53, 279}

\bibitem[\protect\citeauthoryear{{Marley} et~al.,}{{Marley}
  et~al.}{2021}]{Marley2021}
{Marley} M.~S.,  et~al., 2021, \mn@doi [\apj] {10.3847/1538-4357/ac141d}, \href
  {https://ui.adsabs.harvard.edu/abs/2021ApJ...920...85M} {920, 85}

\bibitem[\protect\citeauthoryear{{Marois}, {Macintosh}, {Barman}, {Zuckerman},
  {Song}, {Patience}, {Lafreni{\`e}re}  \& {Doyon}}{{Marois}
  et~al.}{2008}]{Marois2008}
{Marois} C.,  {Macintosh} B.,  {Barman} T.,  {Zuckerman} B.,  {Song} I.,
  {Patience} J.,  {Lafreni{\`e}re} D.,   {Doyon} R.,  2008, \mn@doi [Science]
  {10.1126/science.1166585}, \href
  {http://adsabs.harvard.edu/abs/2008Sci...322.1348M} {322, 1348}

\bibitem[\protect\citeauthoryear{{Mesa} et~al.,}{{Mesa}
  et~al.}{2023}]{Mesa2023}
{Mesa} D.,  et~al., 2023, \mn@doi [\aap] {10.1051/0004-6361/202345865}, \href
  {https://ui.adsabs.harvard.edu/abs/2023A&A...672A..93M} {672, A93}

\bibitem[\protect\citeauthoryear{{Mikhailenko}, {Babikov}  \&
  {Golovko}}{{Mikhailenko} et~al.}{2005}]{Mikhailenko2005}
{Mikhailenko} S.~N.,  {Babikov} Y.~L.,   {Golovko} V.~F.,  2005, Atmospheric
  and oceanic optics, 18, 685

\bibitem[\protect\citeauthoryear{{Miles}, {Skemer}, {Barman}, {Allers}  \&
  {Stone}}{{Miles} et~al.}{2018}]{Miles2018}
{Miles} B.~E.,  {Skemer} A.~J.,  {Barman} T.~S.,  {Allers} K.~N.,   {Stone}
  J.~M.,  2018, \mn@doi [\apj] {10.3847/1538-4357/aae6cd}, \href
  {https://ui.adsabs.harvard.edu/abs/2018ApJ...869...18M} {869, 18}

\bibitem[\protect\citeauthoryear{{Miles} et~al.,}{{Miles}
  et~al.}{2023}]{Miles2023}
{Miles} B.~E.,  et~al., 2023, \mn@doi [\apjl] {10.3847/2041-8213/acb04a}, \href
  {https://ui.adsabs.harvard.edu/abs/2023ApJ...946L...6M} {946, L6}

\bibitem[\protect\citeauthoryear{{M{\"u}ller} et~al.,}{{M{\"u}ller}
  et~al.}{2018}]{Mueller2018}
{M{\"u}ller} A.,  et~al., 2018, \mn@doi [\aap] {10.1051/0004-6361/201833584},
  \href {https://ui.adsabs.harvard.edu/abs/2018A&A...617L...2M} {617, L2}

\bibitem[\protect\citeauthoryear{{Naud} et~al.,}{{Naud}
  et~al.}{2014}]{Naud2014}
{Naud} M.-E.,  et~al., 2014, \mn@doi [\apj] {10.1088/0004-637X/787/1/5}, \href
  {https://ui.adsabs.harvard.edu/abs/2014ApJ...787....5N} {787, 5}

\bibitem[\protect\citeauthoryear{{Nealon}, {Dipierro}, {Alexander}, {Martin}
  \& {Nixon}}{{Nealon} et~al.}{2018}]{Nealon2018}
{Nealon} R.,  {Dipierro} G.,  {Alexander} R.,  {Martin} R.~G.,   {Nixon} C.,
  2018, \mn@doi [\mnras] {10.1093/mnras/sty2267}, \href
  {https://ui.adsabs.harvard.edu/abs/2018MNRAS.481...20N} {481, 20}

\bibitem[\protect\citeauthoryear{{Notsu}, {Eistrup}, {Walsh}  \&
  {Nomura}}{{Notsu} et~al.}{2020}]{Notsu2020}
{Notsu} S.,  {Eistrup} C.,  {Walsh} C.,   {Nomura} H.,  2020, \mn@doi [\mnras]
  {10.1093/mnras/staa2944}, \href
  {https://ui.adsabs.harvard.edu/abs/2020MNRAS.499.2229N} {499, 2229}

\bibitem[\protect\citeauthoryear{{Partridge} \& {Schwenke}}{{Partridge} \&
  {Schwenke}}{1997}]{Partridge1997}
{Partridge} H.,  {Schwenke} D.~W.,  1997, \mn@doi [\jcp] {10.1063/1.473987},
  \href {https://ui.adsabs.harvard.edu/abs/1997JChPh.106.4618P} {106, 4618}

\bibitem[\protect\citeauthoryear{{Petit dit de la Roche}, {Hoeijmakers}  \&
  {Snellen}}{{Petit dit de la Roche} et~al.}{2018}]{Roche2018}
{Petit dit de la Roche} D.~J.~M.,  {Hoeijmakers} H.~J.,   {Snellen} I.~A.~G.,
  2018, \mn@doi [\aap] {10.1051/0004-6361/201833384}, \href
  {https://ui.adsabs.harvard.edu/abs/2018A&A...616A.146P} {616, A146}

\bibitem[\protect\citeauthoryear{{Petrus} et~al.,}{{Petrus}
  et~al.}{2021}]{Petrus2021}
{Petrus} S.,  et~al., 2021, \mn@doi [\aap] {10.1051/0004-6361/202038914}, \href
  {https://ui.adsabs.harvard.edu/abs/2021A&A...648A..59P} {648, A59}

\bibitem[\protect\citeauthoryear{{Price-Whelan} et~al.,}{{Price-Whelan}
  et~al.}{2018}]{astropy:2018}
{Price-Whelan} A.~M.,  et~al., 2018, \mn@doi [\aj] {10.3847/1538-3881/aabc4f},
  \href {https://ui.adsabs.harvard.edu/#abs/2018AJ....156..123T} {156, 123}

\bibitem[\protect\citeauthoryear{{Ruffio} et~al.,}{{Ruffio}
  et~al.}{2021}]{Ruffio2021}
{Ruffio} J.-B.,  et~al., 2021, \mn@doi [\aj] {10.3847/1538-3881/ac273a}, \href
  {https://ui.adsabs.harvard.edu/abs/2021AJ....162..290R} {162, 290}

\bibitem[\protect\citeauthoryear{{Schneider} \& {Bitsch}}{{Schneider} \&
  {Bitsch}}{2021}]{Schneider2021}
{Schneider} A.~D.,  {Bitsch} B.,  2021, \mn@doi [\aap]
  {10.1051/0004-6361/202039640}, \href
  {https://ui.adsabs.harvard.edu/abs/2021A&A...654A..71S} {654, A71}

\bibitem[\protect\citeauthoryear{{Schneider} et~al.,}{{Schneider}
  et~al.}{2023}]{Schneider2023}
{Schneider} A.~C.,  et~al., 2023, \mn@doi [\apjl] {10.3847/2041-8213/acb0cd},
  \href {https://ui.adsabs.harvard.edu/abs/2023ApJ...943L..16S} {943, L16}

\bibitem[\protect\citeauthoryear{{Sepulveda} \& {Bowler}}{{Sepulveda} \&
  {Bowler}}{2022}]{Sepulveda2022}
{Sepulveda} A.~G.,  {Bowler} B.~P.,  2022, \mn@doi [\aj]
  {10.3847/1538-3881/ac3bb5}, \href
  {https://ui.adsabs.harvard.edu/abs/2022AJ....163...52S} {163, 52}

\bibitem[\protect\citeauthoryear{{Skemer} et~al.,}{{Skemer}
  et~al.}{2016}]{Skemer2016}
{Skemer} A.~J.,  et~al., 2016, \mn@doi [\apj] {10.3847/0004-637X/817/2/166},
  \href {https://ui.adsabs.harvard.edu/abs/2016ApJ...817..166S} {817, 166}

\bibitem[\protect\citeauthoryear{{Spiegel} \& {Burrows}}{{Spiegel} \&
  {Burrows}}{2012}]{Spiegel2012}
{Spiegel} D.~S.,  {Burrows} A.,  2012, \mn@doi [\apj]
  {10.1088/0004-637X/745/2/174}, \href
  {https://ui.adsabs.harvard.edu/abs/2012ApJ...745..174S} {745, 174}

\bibitem[\protect\citeauthoryear{{Tabone} et~al.,}{{Tabone}
  et~al.}{2023}]{Tabone2023}
{Tabone} B.,  et~al., 2023, \mn@doi [arXiv e-prints]
  {10.48550/arXiv.2304.05954}, \href
  {https://ui.adsabs.harvard.edu/abs/2023arXiv230405954T} {p. arXiv:2304.05954}

\bibitem[\protect\citeauthoryear{{Tokunaga} et~al.,}{{Tokunaga}
  et~al.}{1998}]{Tokunaga1998}
{Tokunaga} A.~T.,  et~al., 1998, in {Fowler} A.~M.,  ed.,  Society of
  Photo-Optical Instrumentation Engineers (SPIE) Conference Series Vol. 3354,
  Infrared Astronomical Instrumentation. pp 512--524,
  \mn@doi{10.1117/12.317277}

\bibitem[\protect\citeauthoryear{{Uyama} et~al.,}{{Uyama}
  et~al.}{2021}]{Uyama2021}
{Uyama} T.,  et~al., 2021, \mn@doi [\aj] {10.3847/1538-3881/ac2739}, \href
  {https://ui.adsabs.harvard.edu/abs/2021AJ....162..214U} {162, 214}

\bibitem[\protect\citeauthoryear{{Vos}, {Faherty}, {Gagn{\'e}}, {Marley},
  {Metchev}, {Gizis}, {Rice}  \& {Cruz}}{{Vos} et~al.}{2022}]{Vos2022}
{Vos} J.~M.,  {Faherty} J.~K.,  {Gagn{\'e}} J.,  {Marley} M.,  {Metchev} S.,
  {Gizis} J.,  {Rice} E.~L.,   {Cruz} K.,  2022, \mn@doi [\apj]
  {10.3847/1538-4357/ac4502}, \href
  {https://ui.adsabs.harvard.edu/abs/2022ApJ...924...68V} {924, 68}

\bibitem[\protect\citeauthoryear{{Wallace}, {Ireland}  \&
  {Federrath}}{{Wallace} et~al.}{2021}]{Wallace2021}
{Wallace} A.~L.,  {Ireland} M.~J.,   {Federrath} C.,  2021, arXiv e-prints,
  \href {https://ui.adsabs.harvard.edu/abs/2021arXiv210111130W} {p.
  arXiv:2101.11130}

\bibitem[\protect\citeauthoryear{{Wang} et~al.,}{{Wang}
  et~al.}{2021}]{Wang2021}
{Wang} J.~J.,  et~al., 2021, \mn@doi [\aj] {10.3847/1538-3881/abdb2d}, \href
  {https://ui.adsabs.harvard.edu/abs/2021AJ....161..148W} {161, 148}

\bibitem[\protect\citeauthoryear{{Wilcomb}, {Konopacky}, {Barman}, {Theissen},
  {Ruffio}, {Brock}, {Macintosh}  \& {Marois}}{{Wilcomb}
  et~al.}{2020}]{Wilcomb2020}
{Wilcomb} K.~K.,  {Konopacky} Q.~M.,  {Barman} T.~S.,  {Theissen} C.~A.,
  {Ruffio} J.-B.,  {Brock} L.,  {Macintosh} B.,   {Marois} C.,  2020, \mn@doi
  [\aj] {10.3847/1538-3881/abb9b1}, \href
  {https://ui.adsabs.harvard.edu/abs/2020AJ....160..207W} {160, 207}

\bibitem[\protect\citeauthoryear{{Woitke} \& {Helling}}{{Woitke} \&
  {Helling}}{2004}]{Woitke2004}
{Woitke} P.,  {Helling} C.,  2004, \mn@doi [\aap] {10.1051/0004-6361:20031605},
  \href {https://ui.adsabs.harvard.edu/abs/2004A&A...414..335W} {414, 335}

\bibitem[\protect\citeauthoryear{{Wu} et~al.,}{{Wu} et~al.}{2020}]{Wu2020}
{Wu} Y.-L.,  et~al., 2020, \mn@doi [\aj] {10.3847/1538-3881/ab818c}, \href
  {https://ui.adsabs.harvard.edu/abs/2020AJ....159..229W} {159, 229}

\bibitem[\protect\citeauthoryear{{Zhang}, {Bosman}  \& {Bergin}}{{Zhang}
  et~al.}{2020}]{Zhang2020}
{Zhang} K.,  {Bosman} A.~D.,   {Bergin} E.~A.,  2020, \mn@doi [\apjl]
  {10.3847/2041-8213/ab77ca}, \href
  {https://ui.adsabs.harvard.edu/abs/2020ApJ...891L..16Z} {891, L16}

\bibitem[\protect\citeauthoryear{{Zhang}, {Liu}, {Claytor}, {Best}, {Dupuy}  \&
  {Siverd}}{{Zhang} et~al.}{2021}]{ZhangZ2021}
{Zhang} Z.,  {Liu} M.~C.,  {Claytor} Z.~R.,  {Best} W. M.~J.,  {Dupuy} T.~J.,
  {Siverd} R.~J.,  2021, \mn@doi [\apjl] {10.3847/2041-8213/ac1123}, \href
  {https://ui.adsabs.harvard.edu/abs/2021ApJ...916L..11Z} {916, L11}

\bibitem[\protect\citeauthoryear{{Zhu}, {Ju}  \& {Stone}}{{Zhu}
  et~al.}{2016}]{Zhu2016}
{Zhu} Z.,  {Ju} W.,   {Stone} J.~M.,  2016, \mn@doi [\apj]
  {10.3847/0004-637X/832/2/193}, \href
  {https://ui.adsabs.harvard.edu/abs/2016ApJ...832..193Z} {832, 193}

\bibitem[\protect\citeauthoryear{{van der Marel} et~al.,}{{van der Marel}
  et~al.}{2018}]{vanderMarel2018}
{van der Marel} N.,  et~al., 2018, \mn@doi [\apj] {10.3847/1538-4357/aaaa6b},
  \href {https://ui.adsabs.harvard.edu/abs/2018ApJ...854..177V} {854, 177}

\makeatother
\end{thebibliography}

\bsp	
\label{lastpage}
\end{document}